\documentclass[a4paper,11pt]{article}
\pdfoutput=1 

\usepackage{jinstpub} 
\usepackage{palatino}
\usepackage{graphicx}
\usepackage{psfig}
\usepackage{subcaption}
\usepackage{enumitem}

\usepackage{lineno}

\title{\boldmath{Investigation and improvements of the mechanical structure
of Cylindrical GEMs for the BESIII experiment}}

\author[a,b,1]{I.~Balossino,\note{Corresponding author.}}

\affiliation[a]{INFN-Ferrara,\\via Saragat 1, 44122 Ferrara, Italy} 
\affiliation[b]{Institute of High Energy Physics, Chinese Academy of Sciences,\\ 19B Yuquan Road, Shijingshan District, 100049 Beijing, China}

\emailAdd{balossino@fe.infn.it}

\abstract{Gas Electron Multipliers (GEMs) can be produced in large foils and molded in different shapes. The possibility to create cylindrical layers has opened the opportunity to use such detector as internal tracker at collider experiments. One crucial item is to have low material budget in the active area, so the supporting structure of anode and cathode must be light. 
KLOE2 collaboration has built the first Cylindrical GEM detector with honeycomb material with carbon fiber skins produced at high temperature.
BESIII is developing an innovative CGEM detector with charge and time readout. Among several innovative features, the mechanical structure was designed to be a sandwich of Rohacell and Kapton, a PMI foam. After the transportation of a first production of the detectors from the construction site in Italy to the Institute of High Energy Physics in Beijing, some malfunctions have been observed in some of them, compatible with GEMs deformation inside the detector.
We have performed a detailed study by means of an industrial CT scan available in IHEP laboratory and autopsy to the damaged detectors. In this talk, we will review the construction process, the shipment, the findings of the investigation.
A new supporting structure of carbon fiber and honeycomb, assembled at room temperature, has been designed and developed. The thickness of the carbon fiber is small enough to keep the material budget of a single detector layer below 0.5$\%$ of a radiation length, while the mechanical robustness results beyond the purpose of a detector for HEP.
A first detector with such a mechanical structure has been built and shipped to IHEP, preliminary results from operation (e.g. current stability, discharges, temperature and humidity correlation) of the detectors are also presented in this paper.}


\collaboration[c]{on behalf of BESIII CGEM-IT collaboration}

\proceeding{International Conference Instrumentation for Colliding Beam Physics\\
  24-28 February 2020\\
  Budker Institute of Nuclear Physics, Novosibirsk, Russia}

\begin{document}
\maketitle

\section{Introduction}
BESIII (BEijing Spectrometer) is an experiment located at the Institute of High Energy Phisics in Beijing, China. It is situated on the interaction point of BEPCII (Beijing Electron Positron Collider), a double-ring e$^+$e$^-$ collider that allows to run with a center of mass energy range from 2 GeV to 4.6 GeV and with a peak luminosity of 10$^{33}$cm$^{-2}$s$^{-2}$. The goal is to deepen the studies of hadron physics and $\tau$-charm physics. From the interaction point, the experiment is composed by a Multilayer Drift Camber (MDC), a Time-Of-Flight (TOF) system, a CsI(Tl) electromagnetic calorimeter (EMC), all enclosed in a superconducting solenoidal magnet providing a 1.0 T magnetic field. The solenoid is supported by an octagonal flux-return yoke with resistive plate counter muon identifier modules interleaved with steel \cite{a}. 

Accelerator and experiment are undergoing an upgrade that will allow to continue the data taking for almost other 10 years. From the machine point of view this involves an increase of the energy of the center of mass up to 4.9 GeV, a top-up injection that increases of about 30$\%$ the integrated luminosity and that allows to acquire data continuously and an improvement of the radio-frequency cavities. In the meantime some of the sub-detectors involved in the experiment improved the setup to have better performance \cite{b}. 

The upgrade related with the activites presented in this document involves the inner layers of the MDC. They are showing aging effects due to the high radiation. The plan is to replace them with a Cylindrical GEM (Gas Electron Multipliers) detector.

The new inner tracker is composed by three concentric independent detectors. Each one of them is composed by five layers: a cathode, three GEM foils and an anode readout~\cite{c}. The mechanical structure is composed by permaglass rings at the end of the cylinders to sustain the foils, operating also as gas sealing structure and gap spacers, together with support material glued to anode and cathode electrodes to provide rigidity. The readout anode circuit is composed by X and V strips with a relative stereo angle that depends on the layer geometry. An innovative readout with a dedicated ASIC has been developed and it provides time and charge information for each strip~\cite{d}. 

This design, that involves a series of innovations and special attributes, allows to cope with the requirements of the experiment. The spatial resolution will be around 130 \textmu m in transverse direction and better than 1 mm along the beam axis; the material budget will be $\leq$ 1.5 $\%$ X$_0$ for the whole system; it will cover the 93 $\%$ of the solid angle and will be able to manage a hit rate up to 10$^4$ Hz/cm$^2$ thanks to its high rate capability. 

The whole detector will be build in Italy and then shipped to Beijing for commissioning and installation. Prototypes for each layer have been built to develop and test all the components and to define the steps needed to bring the detector in China ready for installation.

\section{First Production}

A first production for each layer (inner, middle, outer) that compose the full detector has been made. It has been decided to use a sandwich of a PMI foam (Rohacell) and Kapton as mechanical structure for cathode and anode. This decision was made to guarantee low material budget, together with enough strength to have a maximum mechanical deformation of 230 \textmu m if a force of 10 N is applied transversely on one end of the cylinder while the other end is kept fixed. 

Their characterization was performed step by step to allow the preparation of a \textit{Quality Assurance and Quality Control Protocol}: gas check, capacitance and resistance measurements of the high voltage channels and power-on procedure, readout board installation and check, monitoring system. 

All the detectors have been then shipped to Beijing in three separated custom made boxes. They have been designed to hold the detector by the permaglass rings with an axial support inside connected to the rest of the box with four springs to attenuate the vibrations. They travelled as cargo shipment with a commercial flight. 

Once in Beijing, the Quality Assurance and Quality Control Protocol was implemented and some issues were found in the inner and the outer layers. Some of the high voltage channels were draining more current than expected.

In order to be able to have a deeper understanding of what went wrong, a campaign of investigations started.
Different tests have been performed: 
\begin{itemize}
\item Resistance and Capacitance Measurements. For each high voltage sector of the detectors, measurements of the resistance and the capacitance of and between the foils have been performed. This allowed to identify shortcuts and deformations inside the detector and to have a first map of the situation inside the detector.
 
\item Laser Surface Measurement. The external structure of all the detectors shipped to Beijing has been measured with a laser machine. The goal was to bring to light any defect caused by inappropriate torsion, stretch or compression during the handling or the assembling of the layer together. The measurements were perfomed with a laser tracker (figure~\ref{fig1}) acquiring as many points as possible of the surface to reach the highest precision. The test was perfomed by measuring the circumference in different points along the length of the detector on the part facing the laser pointer. To increase the precision, the same measurements were perfomed multiple times. The results, obtained with a precision of few hundreds \textmu m, did not point out any major issue of the outermost part of the detector. 

\begin{figure}[t!]
\centering
\begin{minipage}[c]{0.45\textwidth}
\centering
    \includegraphics[width=\textwidth]{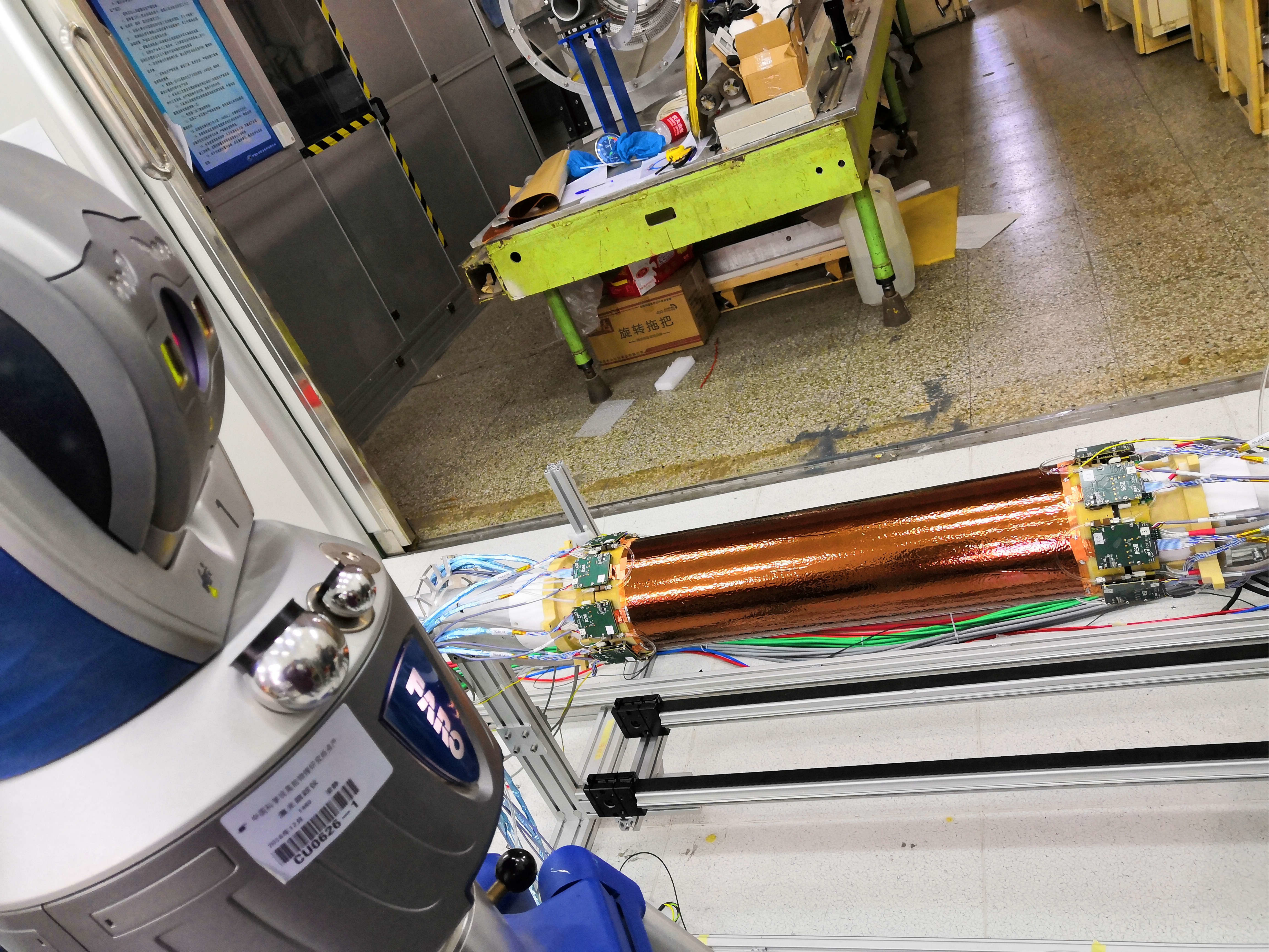}
    \caption{Laser Measurements setup}
    \label{fig1}
\end{minipage}
\begin{minipage}[c]{0.45\textwidth}
\centering
    \includegraphics[width=\textwidth]{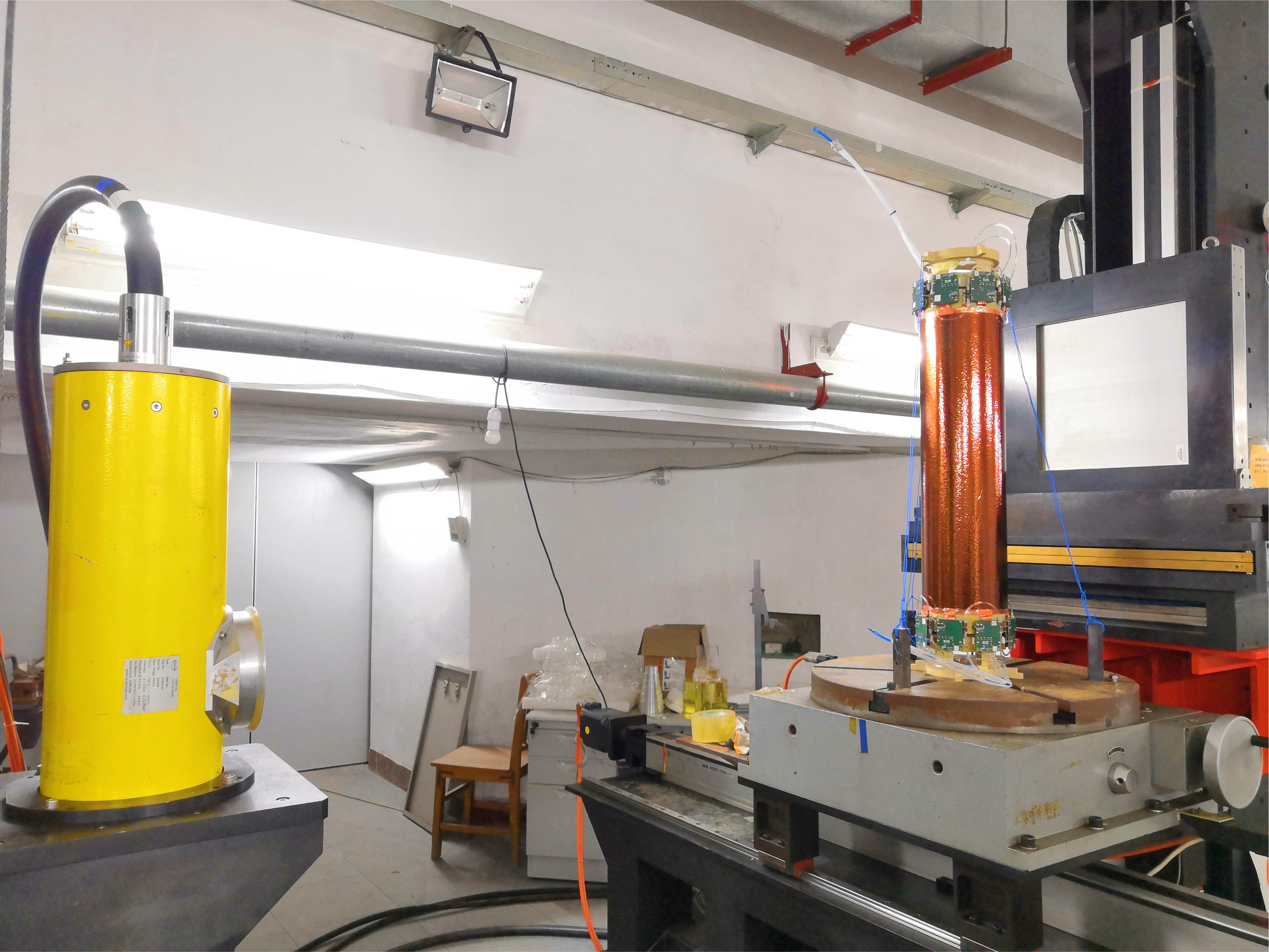}
    \caption{Computed Tomography setup}
    \label{fig2}
\end{minipage}
\end{figure}
\item Computed Tomography Scan. It has been decided to try to have a look inside the detector thanks to a industrial scan (450 keV) available in the same institute of our laboratory. Since it was the first time for such detectors to perform this kind of measurement, it has been decided to test both working and problematic detectors. This could help to run a comparative analysis of any possible issue that we could have found. 

The operation started performing few X-ray scans of the detector in the horizontal position to adjust the right intensity of the beam in order to properly distinguish gaps and foils inside the detector. Then, the detector was placed in vertical position on a rotating platform and properly secured mechanically as presented in figure~\ref{fig2}. With the detector in vertical position, a scan along its length has been performed by extracting different slices of the detector in different measurement points. An example of the resulting images is reported in figure~\ref{fig3}.

\begin{figure}[b!]
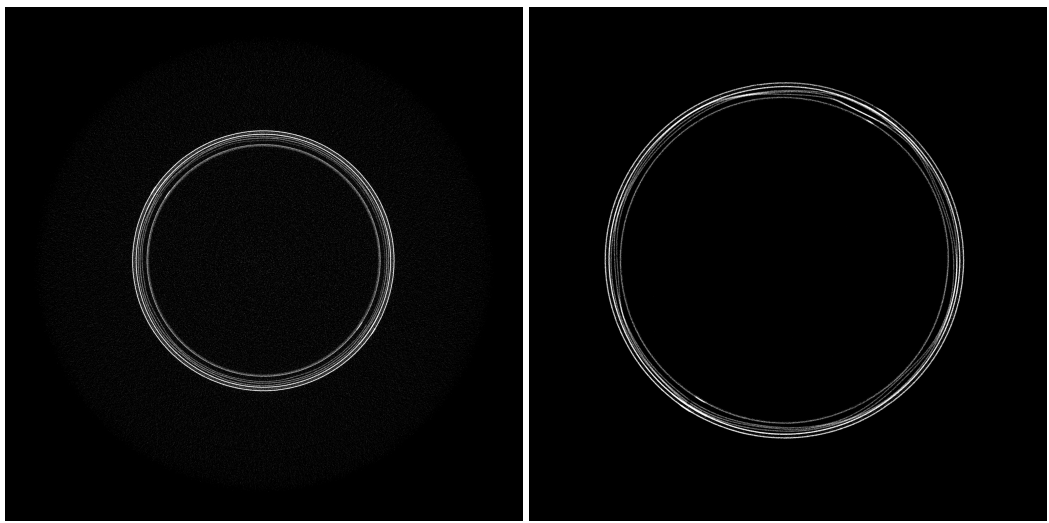

\centering
\begin{subfigure}[c]{0.45\textwidth}
\centering
    \includegraphics[width=\textwidth]{fig3.pdf}
    \caption{Working detector - middle layer}
    \label{fig3a}
\end{subfigure}
\begin{subfigure}[c]{0.45\textwidth}
\centering
    \includegraphics[width=\textwidth]{fig4.pdf}
    \caption{Prolematic detector - outer layer}
    \label{fig3b}
\end{subfigure}
 \caption{CT result of a bad detector}\label{fig3}
\end{figure}

In both figures it is possible to distinguish, from inside to outside, the cathode with the Rohacell-Kapton sandwich, the three GEMs, the Anode, the sandwich of Rohacell-Kapton and the ground plane as most external circle. The resolution of the machine does not allow to picture the exact sizes of the foils and the gap but the important thing is to be able to see any distortion. The difference between the figures is given by the fact that in figure~\ref{fig3a} a 'good detector' is presented, where all the foils are almost equally separated along the circumference, while figure~\ref{fig3b} is a 'bad detector', where it is possible to notice that the GEM foils are touching in more than one point.

The results of such innovative test brought light to what we already found with the measurements of resistance and capacitance: where a shortcut was found, it is possible to see that the foils are touching. 

\item Mechanical Opening. To conclude this mechanic investigation it has been decided to open two of the non-working layers and check them inside. This operation was performed foil by foil, starting from the ground plane, scratching out the Rohacell to remove the mechanical structure, and checking the GEM foils and understand how the defects underlined by the CT presents themselves. By opening the detector, it was also possible to control if issues emerged with the mechanical structure composed by permaglass ring and two sandwiches of Rohacell and Rohacell. No major issues were found but some fragilities were pointed out. The chosen mechanics was enough for this kind of detector and its usual handling and operations. Its fragilities come out under a higher mechanic stress like the one produced by the shipping.

\end{itemize}

The outcome of these investigations brought us to understanding that the mechanical structure chosen for these detectors (Rohacell-Kapton sandwiches) was good for the material budget and for the normal operations, but it is critical for the internal structure of the detector in situations with an excess of vibration or in unexpected events like a turbolence during the shipping from Italy to China. It has been decided to proceed for the final detectors with an improved mechanical robustness. 

\section{Final detectors}

Several tests and simulations have been performed to assure the right rigidity without exceeding the material budget limits defined by the experiment. Different materials have been taken into consideration such as fiberglass or carbon fiber and honeycomb. 

The final detectors present the same design as the prototypes but they will differ for the mechanical support of cathode and anode. Instead of a sandwich of Rohacell-Kapton, a combination of carbon fiber and honeycomb will be used. Simulations calculated that the new structure will sustain a maximum deformation of 15 \textmu m if one end of the cylinder is kept fixed and on the other 10 N are applied transversally while keeping the material budget in the limits of 1.5$\%$ X$_0$.

The innermost layer has been built with this new technology and already shipped to Beijing. Also the shipment was improved in this case by wrapping the whole detector in a foam and secure it in two different boxes to guarantee thermal insulation. 

The detector arrived in Beijing and the Quality Assurance and Quality Control Protocol was applied. No issue was found and the installation of the readout boards continued to be able to assemble it with the middle layer already in Beijing and to start with the data acquisition. 

\acknowledgments

The research leading to these results has been performed within the FEST Project, funded by the European Commission in the call RISE-MSCA-H2020-2020.
The author of this proceeding is supported by the PIFI initiative of the Chinese Academy of Science. 



\begin{thebibliography}{99}

\bibitem{a}
M. Ablikim et al., \emph{Design and construction of the BESIII detector}, NIM A 614 (2010)

\bibitem{b}
M. Ablikim et al., \emph{Future physics program of BESIII}, Chin.Pys.C 44 (2020), doi:10.1088/1674-1137/44/4/040001

\bibitem{c}
R. Farinelli et al., \emph{A cylindrical GEM inner tracker for the BESIII experiment at IHEP}, Springer Proc. Phys. 213 (2018), doi:10.1007/978-981-13-1316-5$\_$21

\bibitem{d}
M. Da Rocha Rolo et al., \emph{A custom readout electronics for the BESIII CGEM detector}, JINST (2017), doi:10.1088/1748-0221/12/07/C07017



\end{thebibliography}
\end{document}